\documentclass[apj]{emulateapj}
\usepackage{graphics,graphicx,times,natbib,longtable,placeins}

\newcommand{\sersic}{{S\'{e}rsic}}
\def\mhalo{${\rm M_{halo}}$}
\def\mstar{${\rm M_{*}}$}
\def\msun{${\rm\,M_\odot}$}

\def\arcsec{$^{\prime\prime}$}

\def\sigmaone{$\Sigma_1$}
\def\dproj{$d_{proj}$}

\def\spose#1{\hbox to 0pt{#1\hss}}
\def\lta{\mathrel{\spose{\lower 3pt\hbox{$\mathchar"218$}}
     \raise 2.0pt\hbox{$\mathchar"13C$}}}

\shorttitle{Quenching of Dwarf Galaxies}
\shortauthors{Guo et al.}

\begin{document}

\title{Implications of Increased Central Mass Surface Densities for the
Quenching of Low-mass Galaxies}

\author{Yicheng Guo\altaffilmark{1}}
\author{Timothy Carleton\altaffilmark{1}}
\author{Eric F. Bell\altaffilmark{2}}
\author{Zhu Chen\altaffilmark{3}}
\author{Avishai Dekel\altaffilmark{4,5}}
\author{S. M. Faber\altaffilmark{6}}
\author{Mauro Giavalisco\altaffilmark{7}}
\author{Dale D. Kocevski\altaffilmark{8}}
\author{Anton M. Koekemoer\altaffilmark{9}}
\author{David C. Koo\altaffilmark{6}}
\author{Peter Kurczynski\altaffilmark{10}}
\author{Seong-Kook Lee\altaffilmark{11}}
\author{F. S. Liu\altaffilmark{12}}
\author{Casey Papovich\altaffilmark{13,14}}
\author{Pablo G. P\'erez-Gonz\'alez\altaffilmark{15,16}}
\altaffiltext{1}{Department of Physics and Astronomy, University of Missouri, Columbia, MO, 65211, USA; {\it guoyic@missouri.edu}}
\altaffiltext{2}{Department of Astronomy, University of Michigan, Ann Arbor, MI, USA}
\altaffiltext{3}{Shanghai Normal University, Shanghai, China}
\altaffiltext{4}{Center for Astrophysics and Planetary Science, Racah Institute of Physics, The Hebrew University, Jerusalem, Israel}
\altaffiltext{5}{SCIPP, University of California, Santa Cruz, CA, USA}
\altaffiltext{6}{UCO/Lick Observatory, Department of Astronomy and Astrophysics, University of California, Santa Cruz, CA, USA}
\altaffiltext{7}{Department of Astronomy, University of Massachusetts, Amherst, MA, USA}
\altaffiltext{8}{Colby College, Waterville, ME, USA}
\altaffiltext{9}{Space Telescope Science Institute, Baltimore, MD, USA}
\altaffiltext{10}{Observational Cosmology Laboratory (Code 665), NASA Goddard Space Flight Center, Greenbelt, MD, 20771, USA}
\altaffiltext{11}{SNU Astronomy Research Center, Department of Physics and Astronomy, Seoul National University, Seoul, Korea}
\altaffiltext{12}{Key Laboratory of Space Astronomy and Technology, National Astronomical Observatories, CAS, Beijing 100101, China}
\altaffiltext{13}{Department of Physics and Astronomy, Texas A\&M University, College Station, TX, USA}
\altaffiltext{14}{George P. and Cynthia Woods Mitchell Institute for Fundamental Physics and Astronomy, Texas A\&M University, College Station, TX, USA}
\altaffiltext{15}{Departamento de Astrof\'{\i}sica, Facultad de CC.  F\'{\i}sicas, Universidad Complutense de Madrid, E-28040 Madrid, Spain}
\altaffiltext{16}{Centro de Astrobiolog\'{\i}a (CAB, INTA-CSIC), Carretera de Ajalvir km 4, E-28850 Torrejón de Ardoz, Madrid, Spain}



\begin{abstract} 

We use the Cosmic Assembly Deep Near-infrared Extragalactic Legacy Survey
(CANDELS) data to study the relationship between quenching and the stellar mass
surface density within the central radius of 1 kpc (\sigmaone) of low-mass
galaxies (stellar mass $M_* \lesssim 10^{9.5} M_\odot$) at $0.5 \leq z < 1.5$.
Our sample is mass complete down to $\sim 10^9$\msun\ at $0.5 \leq z < 1.0$. We
compare the mean \sigmaone\ of star-forming galaxies (SFGs) and quenched
galaxies (QGs) at the same redshift and \mstar. We find that low-mass QGs have
higher \sigmaone\ than low-mass SFGs, similar to galaxies above $10^{10}
M_\odot$. The difference of \sigmaone\ between QGs and SFGs increases slightly
with \mstar\ at $M_* \lesssim 10^{10} M_\odot$ and decreases with \mstar\ at
$M_* \gtrsim 10^{10} M_\odot$. The turnover mass is consistent with the mass
where quenching mechanisms transition from internal to environmental quenching.
At $0.5 \leq z < 1.0$, we find that the \sigmaone\ of galaxies increases by
about 0.25 dex in the green valley (i.e., transitioning region from star
forming to fully quenched), regardless of their \mstar. Using the observed
specific star formation rate (sSFR) gradient in the literature as a constraint,
we estimate that the quenching timescale (i.e., time spent in the transition)
of low-mass galaxies is a few ($\sim4$) Gyrs at $0.5 \leq z < 1.0$. The
responsible quenching mechanisms need to gradually quench star formation in an
outside-in way, i.e., preferentially ceasing star formation in outskirts of
galaxies while maintaining their central star formation to increase \sigmaone.
An interesting and intriguing result is the similarity of the growth of
\sigmaone\ in the green valley between low-mass and massive galaxies, which
suggests that the role of internal processes in quenching low-mass galaxies is
a question worthy of further investigation.

\end{abstract}

\section{Introduction}
\label{intro}

Processes of quenching star formation in galaxies can be divided into two
categories: internal and external. The latter, or environmental quenching, is
believed to be the primary mechanism of quenching galaxies with stellar masses
(\mstar) lower than $10^{9.5}$\msun\ (or low-mass galaxies). Evidence of it
has been obtained at various cosmic epochs
\citep[e.g.,][]{peng10,peng12,geha12,wetzel13,
wheeler14,fillingham15,joshualee15,balogh16,fossati17,ycguo17dwarfenv,nancy17,woo17,ji18,papovich18}.
\citet{peng10} showed that environmental and internal (mass) quenching is
separable in the local universe and the importance of environmental (plus
merger) quenching overtakes that of mass (internal) quenching around
$10^{10}$\msun. \citet[][]{ycguo17dwarfenv} used the projected distance
(\dproj) of a galaxy to its nearest massive neighbor with
\mstar$>10^{10.5}$\msun\ as the indicator of environmental quenching and found
that low-mass quenched galaxies (QGs) have systematically smaller \dproj\ than
star-forming galaxies (SFGs), consistent with the scenario that if
environmental effects are significantly responsible for quenching low-mass
galaxies, QGs should live preferentially within a massive dark matter halo and
hence close to a massive central galaxy. A question, however, remains to be
more thoroughly studied: what is the role that internal processes play in the
quenching of low-mass galaxies?

To answer this question, an effective indicator of internal processes is
needed. Studies of massive galaxies suggest that central mass surface density
within the central radius of 1 kpc (\sigmaone) is such an indicator.
\citet{cheung12} found that high \sigmaone\ is the best predictor of quenching
at $z \sim 0.7$.  \citet{fang13} found that for nearby SDSS galaxies, specific
star formation rate (sSFR) varies systematically relative to \sigmaone,
suggesting a mass-dependent threshold of \sigmaone\ for the onset of quenching,
possibly due to a threshold in black hole mass. \citet{vandokkum14},
\citet{tacchella15a}, and \citet{barro17} extended the use of \sigmaone\ as a
quenching indicator to higher redshifts.  Since massive galaxies are primarily
quenched by internal processes
\citep[e.g.][]{darvish15,darvish16,lin16,delucia19}, \sigmaone\ hence can be
treated as an indicator of internal quenching. Moreover, \citet{terrazas16b}
found that quiescence is strongly correlated with black hole mass in central
galaxies, suggesting that \sigmaone\ is very likely a proxy of black hole mass.
\citet{chen20} proposed a model to correlate \sigmaone\ with the mass of
supermassive black holes. Recently, \citet{luo20} correlated \sigmaone\ with
the bulge types in nearby galaxies. 

In this paper, we use \sigmaone\ to explore quenching in low-mass galaxies. 
Comparing the \sigmaone\ of QGs and SFGs would reveal clues on the quenching
mechanisms. \citet{woo17} studied the \sigmaone\ between SFGs and QGs in SDSS
galaxies, and found that in all environments, at a given \mstar, QGs have a
0.2-0.3 dex higher \sigmaone\ than SFGs. They argued that either \sigmaone\
increases subsequent to satellite quenching, or the \sigmaone\ for individual
galaxies remains unchanged, but the \mstar\ of galaxies at the time of
quenching is significantly different from those in the green valley. Their
sample is down to \mstar=$10^{9.75}$\msun, just around the \mstar, where the
quenching mechanisms are believed to change. Currently, however, similar
studies are lacking for low-mass galaxies beyond the local universe. We will
use the CANDELS data to explore lower-mass regime at $z \geq 0.5$.

The physical meaning of \sigmaone\ in low-mass and massive galaxies, however,
is different. The choice of 1 kpc, which is limited by the spatial resolution
of HST/WFC3, is reasonably small for massive galaxies to sample their central
region, but it is quite close to the effective radius of low-mass galaxies,
raising a concern of its validity to represent the central region. Galaxies,
however, extend well beyond their half-light radius. \citet{mowla19} showed
that the 80\%-light radius ($r_{80}$, i.e., the radius containing 80\% of the
total galaxy light) of low-mass galaxies at $z \lesssim 1.5$ is about 5 kpc,
and this radius increases insignificantly from $10^9$\msun\ to
$10^{10.5}$\msun.  \citet{miller19} showed that the difference in $r_{80}$
between SFGs and QGs is small. Based on these results, we presume that,
although \sigmaone\ might not represent the core density of low-mass galaxies
as it does for massive galaxies, it is still a very good indicator of the mass
surface density in the inner region of galaxies below $10^{10.5}$\msun.
Therefore, as long as we compare galaxies with similar total \mstar, using
\sigmaone\ to study the change of inner mass density before and after quenching
is valid and in fact very important to connect the density change to quenching.

In this paper, we adopt a flat ${\rm \Lambda CDM}$ cosmology with
$\Omega_m=0.3$, $\Omega_{\Lambda}=0.7$, and the Hubble constant $H_0 = 70\ {\rm
km~s^{-1}~Mpc^{-1}}$. We use the AB magnitude scale \citep{oke74} and a
\citet{chabrier03} initial mass function.

\section{Data, Sample, and \sigmaone} \label{data}

We identify QGs and SFGs from CANDELS \citep{candelsoverview,candelshst}
through sSFR. We start from the multi-wavelength photometry catalogs of four
CANDELS fields: COSMOS \citep{nayyeri17cosmos}, EGS \citep{stefanon17egs},
GOODS-S \citep{ycguo13goodss}, and UDS \citep{galametz13uds}.  Photometric
redshifts and \mstar\ were measured by \citet{dahlen13} and \citet{santini15}.
Basically, for each galaxy, 10 investigators ran their spectral energy
distribution (SED)-fitting codes on the galaxy \citep[see Table 1
of][]{santini15}. They were free to choose their preferred star formation
histories ($\tau$-model, constant, delayed, inverse $\tau$, etc.) and many
other free parameters. Nine of them used Calzetti law \citep{calzetti00} to
account for dust extinction. Each investigator delivered the best-fit
parameters (including \mstar\ and star formation rate [SFR]).
We then took the median of the 10 results as the ``team consensus'' parameters
of the galaxy. We believe that this method, used in many CANDELS papers, is
able to reduce the uncertainties and systematics of each individual code or
parameter choices used in SED fitting.

%

\sigmaone\ used in this paper is the same as that used in \citet{chen20}.  It
is calculated in three steps. First, we run FAST \citep{kriek09fast} on
multi-aperture photometry of CANDELS galaxies \citep{liufs16,liufs18} to
measure \mstar\ within each aperture, and then derive the mass density within
the central 1 kpc. This \sigmaone\ is the observed (point spread function
(PSF)-smoothed) \sigmaone. Because of the PSF effect, this observed \sigmaone\
is smaller than the intrinsic one.  Next, we calculate a model \sigmaone\ by
using the \sersic\ model of each galaxy \citep{vanderwel12}. We measure the
model H-band band flux within 1 kpc and convert it into \mstar\ with the best
SED-fit mass-to-light ratio. This \sigmaone\ is the model \sigmaone. Last, for
galaxies at a given redshift, we derive an average relation between the
difference in the observed and model \sigmaone\ (defined as the observed minus
model) as a function of the \sersic\ index ($n$). This relation reveals the
average mass difference caused by the PSF effect, which is then added to the
observed \sigmaone\ at a given redshift and $n$. This hybrid method largely
corrects for PSF effects and central deviation from \sersic\ models, providing
a reliable measurement of the intrinsic \sigmaone. {The magnitude of our PSF
correction and its effect on the observed results of \sigmaone\ are shown in
Appendix A.} Our \sigmaone\ agrees excellently with others in the literature
\citep[e.g.,][]{barro17}. 

Our \sigmaone\ measurement is primarily based on aperture photometry, and the
use of \sersic\ models is only to correct for the PSF effect. The accuracy of
the \sersic\ model plays a role in our derived correction and hence in our
measurement. The correction is derived from the average difference of a
population of galaxies. Therefore, for a population of galaxies, as long as
their average \sersic\ measurement is not biased by systematic errors, our
method is reasonable to recover their average (mean or median) \sigmaone. In
our later analysis, we only use the mean or median \sigmaone\ of a given
population to draw our conclusion. \citet{vanderwel12} showed that for
low-\sersic\ index galaxies ($n < 3$) or for small galaxies (effective radius
$r_e < $0.3\arcsec), the systematic error of the \sersic\ index measured from
CANDLES F160W is almost zero (a relative error of 0.02 for $n < 3$ and 0.00 for
$r_e < $0.3\arcsec) down to an F160W magnitude of 25 AB (see their Table 3).
Low-mass galaxies in our study are indeed galaxies with low $n$ or small $r_e$.
Therefore, based on our scientific objective (i.e., understanding the average
behavior of quenched low-mass galaxies), we are able to use the CANDELS F160W
\sersic\ measurements to F160W mag = 25 AB. 

\begin{figure}[htbp] 
\hspace*{-0.5cm}
\includegraphics[scale=0.23, angle=0]{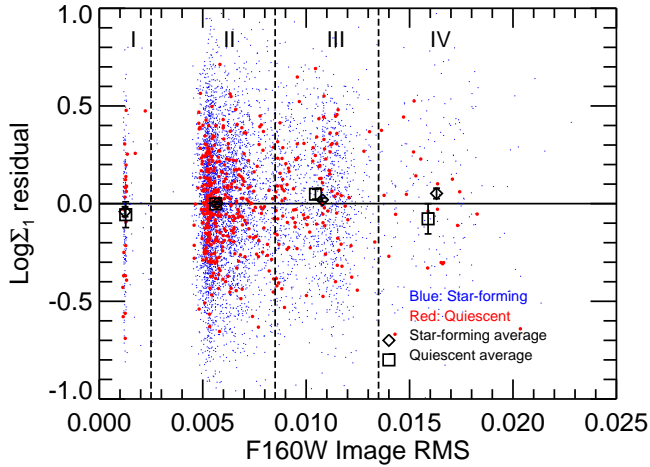}

\caption[]{Comparison of \sigmaone\ in regions with different depths (indicated
by F160W image's root-mean square error RMS) in CANDELS/GOODS-S. Galaxies are
divided into star forming (blue) and quenched (red). For each galaxy, its
\sigmaone\ is compared to the median of galaxies in the Deep region (Region II)
with the same redshift, \mstar, and star formation activity (star forming or
quenched). The difference (individual minus the median) is the \sigmaone\
deviation in the Y-axis. Squares (and diamonds) with error bars show the
3$\sigma$ clipped mean and standard deviation of the mean for quenched (and
star-forming) galaxies. The four regions are: the Hubble Ultra Deep Field
(HUDF, I), Deep (II), Wide (III), and Edge (IV) based on the RMS value of the
F160W image at the location of the galaxies. The smaller the RMS, the deeper
the survey.

\label{fig:hudftest}} \end{figure}

We further test the effect of survey depth on the accuracy of \sigmaone. We
separate galaxies in GOODS-S into regions of the Hubble Ultra Deep Field
(HUDF), Deep, Wide, and Edge (which is for edge or corner regions where the
depth is very shallow due to the drizzle pattern). The survey depth varies
across a range of more than 2.5 mag from the HUDF to the Edge. We choose the
Deep region as our fiducial region due to its decent depth and large number of
galaxies. For each galaxy in any region, we compare its \sigmaone\ to the
median \sigmaone\ of galaxies in the Deep region with the same redshift,
\mstar, and star formation activity (star forming or quenched). The difference
(or \sigmaone\ deviation) is shown in Figure \ref{fig:hudftest}. We find no
obvious difference between the four regions (the largest deviation is that QGs
in the Edge region is about 0.07 dex lower than the other three regions).
Moreover, we find no obvious trend of the deviation changes with the survey
depth. This test provides strong evidence that if limited to F160W = 25 AB,
even though most of CANDELS regions are shallower than HUDF, our \sigmaone\
measurement has little systematic biases to a population of galaxies regardless
of survey depth. {The result in Figure \ref{fig:hudftest} shows the statistics
of the whole sample. To specifically test the robustness of our \sigmaone\
measurement in low-mass galaxies, we repeat it for a subsample of low-mass
galaxies with ${\rm 10^{8.5} < M_*/M_\odot < 10^{9.5}}$ at $0.5<z<1.0$. The new
result, shown in Appendix A, indicates again no significant systematic
deviations found between QGs and SFGs across the four survey depths.}

\begin{figure}[htbp] \includegraphics[scale=0.15,
angle=0]{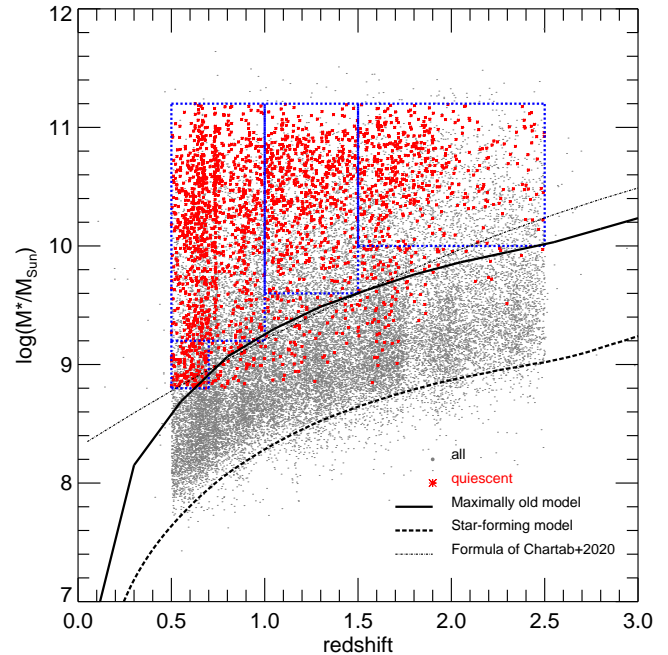}

\caption[]{Sample selection. CANDELS galaxies of $0.5 \leq z < 2.5$ with F160W
H $\leq$ 25 AB are shown as gray points. QGs between $10^{8.8}$ and
$10^{11.2}$\msun\ are red. The three curves show different estimations of mass
completeness of H $\leq$ 25 AB as described in text. The four blue dotted
rectangles show our mass-complete sample of QGs at different redshifts.

\label{fig:sample}} \end{figure}

We therefore limit our low-mass sample to sources with F160W $H\leq25$ AB, no
suspicious F160W photometry, SExtractor CLASS\_STAR$<$0.8, and a clean \sersic\
fit flag in \citet{vanderwel12}. To study its corresponding mass limit, we use
two stellar population models. The first one represents young galaxies with a
constant star formation history and an age of 0.5 Gyr. The second represents
old galaxies with a single starburst right after the big bang. This model has
the maximum age that a galaxy can have at a given redshift. We then normalize
the two models to have the F160W magnitude of 25 AB. The old model (with a high
mass-to-light ratio) corresponds to a much higher stellar mass than the young
one (Figure \ref{fig:sample}). We therefore choose the old one as our mass
limit at different redshifts. As a sanity check, we also use the empirical
formula derived by \citet{chartab20} to measure the mass-complete limit of
F160W = 25 AB \footnote{The formula of mass limit in \citet{chartab20} is for
F160W = 26 AB. We renormalize their formula to match F160W = 25 AB.}. Their
formula is derived by following the method of \citet{pozzetti10}, which is a
common practice in measuring stellar mass functions for large surveys. The
empirical limit matches very well with our model limit at $0.5 \leq z < 2.5$.
We therefore determine that our mass limit is $10^{9.2}$\msun\ at $0.5 \leq z
<1.0$, $10^{9.6}$\msun\ at $1.0 \leq z <1.5$, and $10^{10.0}$\msun\ at $1.5
\leq z < 2.5$ (the three large dotted rectangles in Figure \ref{fig:sample}).
To push our mass limit even lower, we also include galaxies with \mstar\
between $10^{8.8}$\msun\ and $10^{9.2}$\msun\ at $0.5 \leq z < 0.7$, where our
sample is also mass complete (the smallest dotted rectangle in Figure
\ref{fig:sample}).

\begin{figure*}[htbp] 
\includegraphics[scale=0.23, angle=0]{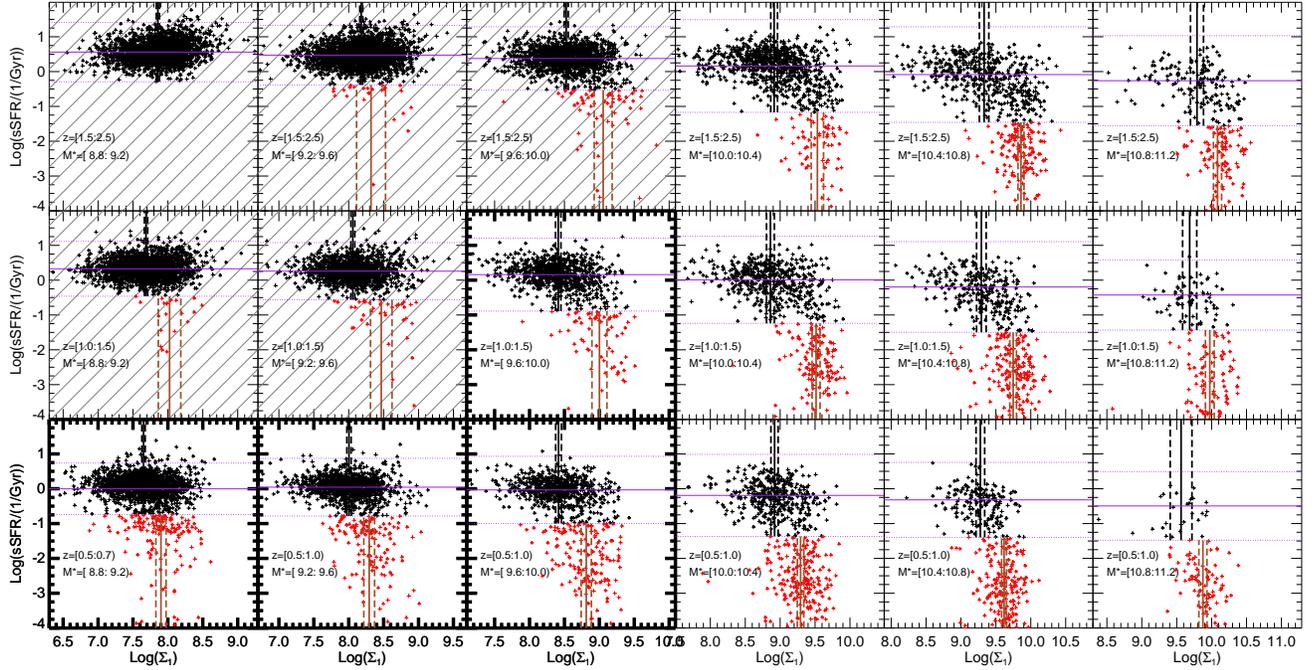}

\caption[]{sSFR as a function of \sigmaone\ in different redshift and \mstar\
bins. In each bin, galaxies are divided into SFGs (black) and QGs (red). The
horizontal solid and dotted purple lines are the 3$\sigma$ clipped mean sSFR
and its 3$\sigma$ deviation for galaxies in the bin. The vertical solid and
dashed black (and brown) lines are the 3$\sigma$ clipped mean \sigmaone\ and
the standard deviation of the mean of \sigmaone\ of SFGs (and QGs). The QG
samples in shaded panels are mass incomplete (SFGs are mass complete, though).
The four panels toward the bottom left with thick borders highlight the main
result of this paper.

\label{fig:ssfrvssigma1}} 
\end{figure*}

\section{Results} \label{results}

\subsection{Difference of \sigmaone\ between QGs and SFGs} \label{results:diff}

Figure \ref{fig:ssfrvssigma1} shows the sSFR as a function of \sigmaone\ in
different redshift and \mstar\ bins. In each bin, first, for galaxies with sSFR
$> 10^{-2} / Gyr$, we calculate the 3$\sigma$ clipped mean and standard
deviation of the mean of sSFRs (purple solid and dotted lines). Galaxies with
an sSFR lower than $-3\sigma$ (the lower dotted line) are classified as QGs,
while the others are classified as SFGs. Our mean sSFR and its scatter
($\sigma$) are consistent with those in the literature
\citep[e.g.,][]{whitaker14,kurczynski16}.

\begin{figure}[htbp] 
\hspace*{-0.3cm}\includegraphics[scale=0.12,
angle=0]{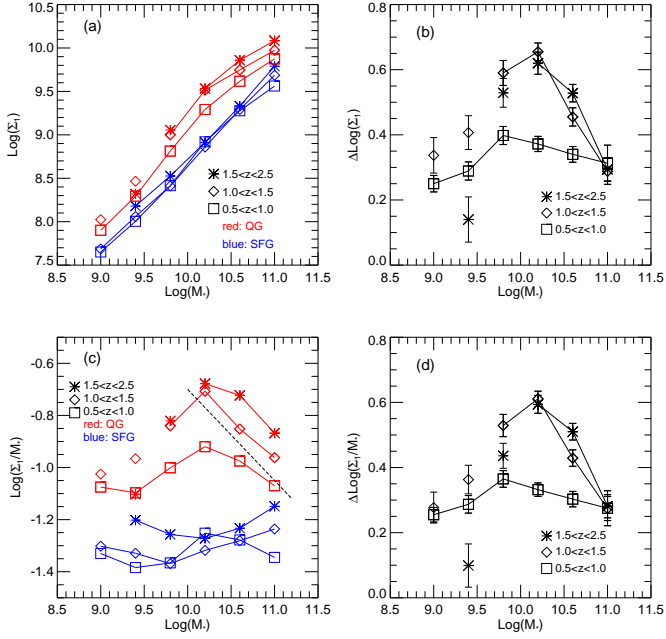}

\caption[]{Statistics of \sigmaone. In each panel, different redshifts are
shown by different symbols, and different types of galaxies are shown by
different colors. Panel (a) shows the 3$\sigma$ clipped mean of \sigmaone\ as a
function of \mstar. Panel (b) shows the difference between the 3$\sigma$
clipped mean of \sigmaone\ of SFGs and QGs ($\Delta$ = QG - SFG). Panel (c)
shows the mass-normalized \sigmaone\ as a function of \mstar. The dashed line
shows the slope derived from \citet{barro17}. Panel (d) is similar to Panel
(b), but showing the difference between the mass-normalized \sigmaone\ of SFGs
and QGs. In each Panel, symbols connected by lines are mass complete in our
sample, while those not connected are mass incomplete.

\label{fig:stat}} 
\end{figure}

Figure \ref{fig:ssfrvssigma1} shows that, for almost all redshift and \mstar\
bins, the 3$\sigma$ clipped mean \sigmaone\ of QGs (solid vertical brown lines)
is greater than that of SFGs (solid vertical black lines). Figure
\ref{fig:stat} shows the statistics. 

For massive galaxies (\mstar$\geq 10^{10}$\msun), at all redshifts, the mean
\sigmaone\ of QGs is significantly ($\gtrsim 10 \sigma$) larger than that of
SFGs with the same redshift and \mstar\ (Panel (b) of Figure \ref{fig:stat}).
The difference decreases with \mstar. This result is consistent with many other
studies at various redshifts
\citep[e.g.,][]{cheung12,fang13,vandokkum14,barro17,chen20}. This difference in
\sigmaone, combined with no difference in \dproj\ between massive QGs and SFGs
found in \citet{ycguo17dwarfenv}, demonstrates that internal processes,
especially those related to central mass density, e.g., possibly black holes,
are the primary way to quench massive galaxies from $z=2.5$ to $z=0.5$
\citep[e.g.,][]{chen20} or even to today \citep[e.g.,][]{luo20}.

The new, interesting result is the \sigmaone\ comparison in the low-mass regime
(i.e., panels highlighted by thick borders in Figure \ref{fig:ssfrvssigma1}).
At \mstar$\lesssim 10^{9.5}$\msun\, we find a $\gtrsim$0.3 dex difference with
very high statistical significance ($\gtrsim 5\sigma$) between the mean
\sigmaone\ of QGs and SFGs at $z<1.5$. This result suggests that quenching has
a relation with the increase of \sigmaone\ in low-mass galaxies, similar to the
situation of massive galaxies reported by many studies. We will discuss its
implications in detail in Section \ref{discussion}.

To remove the mass dependence of \sigmaone\ further, we calculate the ratio
between \sigmaone\ and \mstar\ of galaxies (called mass-normalized \sigmaone,
which is proportional to the fraction of total \mstar\ within the central 1kpc
radius, but differs by a factor of $\pi$). As shown in Panel (c) of Figure
\ref{fig:stat}, this mass-normalized \sigmaone\ decreases with \mstar\ for
massive QGs. Our trend matches the slope (dashed line) derived through the
best-fit \sigmaone--\mstar\ relation of QGs from \citet{barro17}. For low-mass
galaxies at $z<1.0$, our QG trend shows a turnover at $10^{10}$\msun, and
decreases with the decreasing of \mstar. This turnover is mainly a reflection
of changes in the \sersic\ index ($n$) and size (see Figure 13 of
\citet{barro17}). For $n \sim 4$ galaxies, the mass-normalized \sigmaone\
decreases when galaxy size increases. Massive QGs have $n \sim 4$ and their
size increases with \mstar. Therefore, their mass-normalized \sigmaone\
decreases with \mstar. However, for low-mass QGs at $z \leq 1.0$, their size is
flattened in the \mstar--size diagram \citep[e.g.,][]{vanderwel14size}.
Moreover, their $n$ gradually changes from $\sim 4$ to $\sim 1$ with the
decrease of \mstar. The combination of the two (constant size and decreasing
$n$) makes their mass-normalized \sigmaone\ decreases moderately from the peak
at $10^{10}$\msun\ and approach a constant value of -1.1 dex. This turnover at
$10^{10}$\msun\ is also around the stellar mass where quenching mechanisms
gradually changes from being dominated by internal processes to by external
processes
\cite[e.g.,][]{peng10,cybulski14,joshualee15,darvish15,darvish16,wetzel15a,balogh16,fillingham16,lin16,ycguo17dwarfenv,lee18}.

\begin{figure}[htbp]
\hspace*{-0.3cm}\includegraphics[scale=0.12,
angle=0]{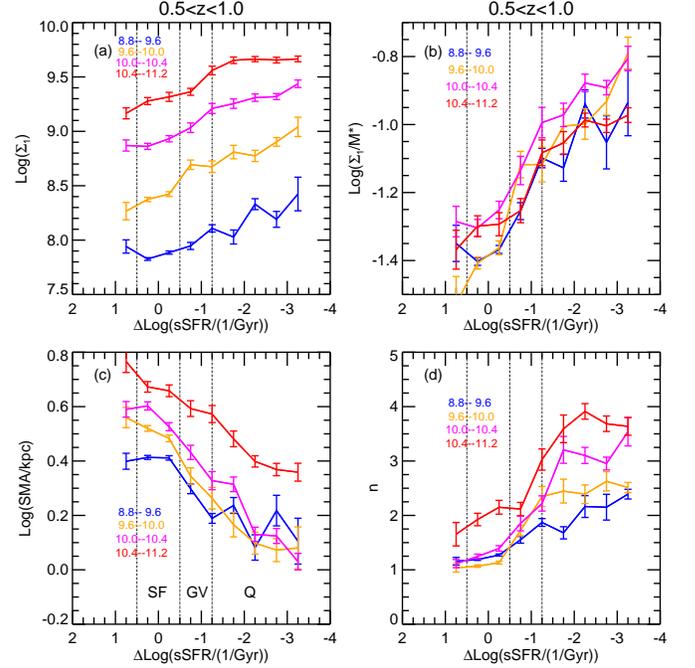}

\caption[]{Relations between different parameters and the residual Log(sSFR) of
galaxies at $0.5 \leq z < 1.0$. The residual Log(sSFR), or $\Delta$Log(sSFR),
is defined as the difference between the Log(sSFR) of a galaxy and the average
Log(sSFR) of galaxies on the star-forming main sequence at the same redshift
and \mstar\ of the galaxy. Namely, the zero-point of $\Delta$Log(sSFR)
corresponds to the solid horizontal purple lines in Figure
\ref{fig:ssfrvssigma1}. The four panels show (a) \sigmaone, (b) mass-normalized
\sigmaone\ (i.e., Log(\sigmaone/$M_*$), (c) semi-major axis, and (d) \sersic\
index. In each panel, galaxies are divided into four \mstar\ bins, as the
labels of different colors show. The curves and error bars show the mean and
the standard deviation of the mean (both after 3$\sigma$ clipping) of each
\mstar\ bin. The three vertical black lines show the boundaries of three
regions, as indicated in Panel (c): star forming (SF), green valley (GV), and
quenched (Q).

\label{fig:lowzstat}}
\end{figure}

\begin{figure}[htbp]
\hspace*{-0.3cm}\includegraphics[scale=0.12,
angle=0]{ssfrvssigma1_stack_25.0_lowz3.ps}

\caption[]{Similar to Figure \ref{fig:lowzstat_delta}, but using the residual
values (rather than their exact values) for the parameters in the Y-axis. The
residual of a parameter is defined as the difference between the parameter of a
galaxy and the average of galaxies on the star-forming main sequence at the
same redshift and \mstar\ of the galaxy. For example, $\Delta$Log(\sigmaone) is
defined by using the solid vertical black lines in Figure
\ref{fig:ssfrvssigma1} as the zero-point.

\label{fig:lowzstat_delta}}
\end{figure}

The mass-normalized \sigmaone\ of SFGs is almost a constant across a wide range
of \mstar, and is lower than that of their QG counterparts. The difference of
mass-normalized \sigmaone\ between the two populations (Panel (d) of Figure
\ref{fig:stat}) is very similar to that of \sigmaone\ (Panel (b)). The trend of
the mass-normalized \sigmaone\ of massive SFGs is still uncertain in the
literature. For example, \citet{barro17} found a sublinear slope (0.88) for the
\sigmaone--\mstar\ relation, which turns into a mass-normalized \sigmaone\
slightly decreasing with \mstar. \citet{chen20} revisited this issue and found
a superlinear slope (1.1) for the \sigmaone--\mstar\ relation, which implies
the mass-normalized \sigmaone\ slightly increases with \mstar. Overall, both
slopes are quite close to 1, implying a close-to-flat trend of mass-normalized
\sigmaone. Therefore, our result of an almost constant trend does not
significantly different from either slope.

\subsection{\sigmaone\ and Structural Parameters vs. sSFR} \label{results:delta}

An efficient way to investigate the overall behavior of galaxies over a wide
range of parameter space is to use the ``residual'' plot
\citep[e.g.,][]{barro17,woo17,fang18,chen20,lin20}. Instead of showing the
exact values of a parameter, this type of plot shows the relative values of the
parameter with respect to a common value, for example, the mean or median of a
sample. By taking into account of the different mean or median values of
different samples, this method enables a direct comparison between them. 

We use the residual plot to show the relations between some parameters and the
sSFR. In Figure \ref{fig:lowzstat}, we only use the residual for the sSFR in
the X-axis, but still keep the exact values of other parameters in the Y-axis
for readers who are interested in the exact values. In Figure
\ref{fig:lowzstat_delta}, we use the residual for both axes to compare
different subsamples with different \mstar. The zero-point of the residual is
calculated by the average value of galaxies in the star-forming main sequence
at a given ($z$, \mstar) bin. For example, the zero-point of  $\Delta$Log(sSFR)
(or $\Delta$Log(\sigmaone)) is the solid horizontal purple lines (or solid
vertical black lines) in Figure \ref{fig:ssfrvssigma1}. Hereinafter, we only
focus on the redshift range of $0.5 \leq z < 1.0$. We also divide galaxies into
three categories based on their \mstar: low-mass (Log(\mstar)$<9.6$),
intermediate-mass ($9.6 \leq $Log(\mstar)$<10.4$), and massive ($10.4
\leq$Log(\mstar)$< 11.2$).

To catch the transition between SFGs and QGs, we divide galaxies into three
(rather than previous two) types based on their $\Delta$Log(sSFR): star forming
(SF), green valley (GV\footnote{To be accurate, this type should be called
``transition''. We use ``GV'' simply to follow the naming scheme in studies of
massive galaxies.  There is no evidence yet to show there is a ``valley'' in
between SFGs and QGs in the distributions of low-mass galaxies at $z>0.5$.}),
and quenched (Q). The divisions are shown in Panel (c) of Figure
\ref{fig:lowzstat} and \ref{fig:lowzstat_delta}. They are quite similar to
those used in \citet{woo17,liufs18,chen20}. In SF, \sigmaone\ and
\sigmaone/\mstar of galaxies of all \mstar\ changes little (Panel (a) and (b)
of both figures). In GV, all galaxies increases their \sigmaone\ (or
Log(\sigmaone/\mstar)) dramatically by about 0.2-0.3 dex. The increase of
\sigmaone\ of massive galaxies during GV has been discussed in \citet[][see
their Figure 5]{chen20}.  Our result shows that intermediate- and low-mass
galaxies also follow a similar trend during GV. In Q, galaxies with different
\mstar\ show a slight difference. Massive galaxies reach a plateau of about 0.3
dex in $\Delta$Log(\sigmaone) or $\Delta$Log($\Sigma_1/M_*$). This plateau is a
reflection of the vertical part of the ``elbow'' shape in the panels of massive
galaxies in Figure \ref{fig:ssfrvssigma1}. This trend and its meaning on
quenching have been thoroughly discussed by \citet{barro17}, \citet{woo17}, and
\citet{chen20}. On the other hand, intermediate- and low-mass galaxies continue
to grow their \sigmaone after GV. This trend is especially true for
intermediate-mass galaxies, and less obvious for low-mass galaxies due to large
uncertainties of small number statistics.

The relation between Log(\sigmaone) (or $\Delta$Log(\sigmaone)) and
$\Delta$Log(sSFR) is correlated with the change of other structural parameters
of galaxies. Panel (c) and (d) of Figure \ref{fig:lowzstat} and
\ref{fig:lowzstat_delta} show the change of size (semimajor axis) and the
\sersic\ index ($n$). In GV and Q, the sizes of galaxies decrease with
$\Delta$Log(sSFR). Low-mass and massive galaxies have a similar decreasing
trend, while intermediate-mass galaxies decrease faster. Although
intermediate-mass galaxies are larger than low-mass ones in SF, because of this
faster decrease, their size approach the size of low-mass galaxies when
$\Delta$Log(sSFR) is very low (Panel (c) of Figure \ref{fig:lowzstat}). This
result is consistent with the flattened curve of the size--mass relation of QGs
in the low-mass regime \citep[e.g.,][]{vanderwel14size}.

Low-mass galaxies and massive galaxies, however, have different trends of $n$.
Low-mass galaxies start with a disk-like shape ($n=1$) in SF, then mildly
increase $n$ in GV and Q, and finally to reach $n \sim 2$ -- still disk-like --
in the end of Q. Massive galaxies already start with $n \sim 2$ in SF and
quickly increase $n$ during GV and the beginning of Q to $n \sim 4$
(spheroid-like) and remain around $n \sim 4$ in Q. Intermediate-mass galaxies
lie between the two extremes. Overall, to summarize, massive galaxies' plateau
of Log(\sigmaone) at $\Delta$Log(sSFR)$<-2$ correlates with (almost) no change
in size and $n$ in Q. Low-mass galaxies' increasing \sigmaone\ in Q correlates
with a gradual increase of $n$ more than with a slight decrease of size.
Intermediate-mass galaxies' increasing \sigmaone\ in Q, on the other hand,
correlates more with a dramatic decrease of size.

\section{Discussion} \label{discussion}

The main result of this study is the difference between the mean \sigmaone\ of
QGs and SFGs in the low-mass regime (\mstar$\lesssim10^{9.5}$\msun) at
$z\sim1$. Similar results have also been reported in the local universe for
galaxies with \mstar$>10^{9.75}$\msun\ in \citet{woo17}. Here, we discuss some
implications of this result for the quenching processes of low-mass galaxies.
Although the real quenching processes are very likely to be complicated and
complex, our discussions are still useful to place some constraints on
quenching and environmental effects.

When we discuss \sigmaone\ and quenching, it would be very useful to separate
low-mass central and satellite galaxies \footnote{See, however, a series of
papers by \citet{wangenci18a,wangenci18b,wangenci20}. They showed that the
separation of central and satellite galaxies is not as fundamental as using
galaxy \mstar\ in discussing quenching mechanisms, because the quenching
properties (e.g., bulge-to-total-light ratio, central velocity dispersion, and
prevalence of optical/radio-loud active galactic nuclei (AGN), etc.) of central
and satellite galaxies are quite similar as long as both stellar mass and halo
mass are controlled. An earlier work by \citet{ycguo09cen} also found similar
structural parameter distributions between central and satellite galaxies when
mass and color are controlled.}. As shown in \citet{woo17}, \sigmaone\ of the
transition galaxies (i.e. green-valley galaxies) decreases smoothly with the
environment by as much as 0.2 dex for \mstar=$10^{9.75}-10^{10}$\msun\ galaxies
from the field, large halo-centric distance, to small halo-centric distance in
SDSS. In our sample, however, due to the limitation of using photometric
redshift, galaxy clusters or groups cannot be accurately identified. In
\citet{ycguo17dwarfenv}, we found that the median projected distance
($d_{proj}$) from low-mass ($M_* \lesssim 10^{9.5}$\msun) QGs to their massive
($M_* \gtrsim 10^{10.5}$\msun) neighbors is systematically smaller than that of
low-mass SFGs. Also, \citet{ycguo17dwarfenv} showed that $\sim90\%$ of low-mass
QGs are within 2 times the virial radius of their neighboring massive halos.
These results suggest that (1) on average, low-mass QGs are found closer to
massive central galaxies (i.e., higher density region) than low-mass SFGs and
(2) most of low-mass QGs tend to be satellite galaxies. Therefore, the results
we see in Figure \ref{fig:ssfrvssigma1} -- \ref{fig:lowzstat_delta} are likely
driven by satellite galaxies.

In next two subsections, we interpret the curves in Figure \ref{fig:lowzstat}
and \ref{fig:lowzstat_delta} as evolutionary tracks. This interpretation is not
necessarily true, especially when progenitor bias is considered (Section
\ref{discussion:bias}). It is, however, still reasonable and useful for our
discussion when the change of \mstar\ of galaxies is small, which is likely
true when galaxies quench. The decrease of sSFR results in a moderate increase
of \mstar, which then keeps most galaxies within their initial \mstar\ bins, as
shown in Figure \ref{fig:lowzstat} and \ref{fig:lowzstat_delta}.

\subsection{Low-mass Galaxies in Green Valley} \label{discussion:gv}

\sigmaone\ of low-mass galaxies increases by about 0.25 dex in GV, indicating
that the quenching mechanisms (possibly environmental effects) gradually quench
star formation in an outside-in process, i.e., preferentially removing gas from
outskirts of galaxies while maintaining their central star formation to
increase \sigmaone. We can use the growth of total \mstar\ and \sigmaone\ to
estimate the timescale of galaxies in GV. We will use the sSFR gradient of
low-mass transition galaxies reported in \citet{liufs18} as our constraint.

We consider a galaxy entering GV with \mstar\ of $10^{9.2}$\msun\ at $z<1.0$.
Its sSFR (and hence SFR) can be measured from our sample and is -0.5 dex below
the average Log(sSFR) of star-forming main sequence (as defined in Figure
\ref{fig:lowzstat}). When this galaxy exits GV, its sSFR drops by another -0.75
dex to reach -1.25 dex below the star-forming main sequence. Now, we assume
this galaxy spends $T_{GV}$=1, 2, and 4 Gyr in GV. Its Log(sSFR) drops linearly
with time over this period (i.e., sSFR drops exponentially). Since the change
of sSFR is only -0.75 dex, very small compared to the whole sSFR range from SF
to Q, the exact form of the decreasing is not important.

\begin{figure}[htbp] 
\hspace*{-0.3cm}
\vspace*{-0.3cm}
\includegraphics[scale=0.12,angle=0]{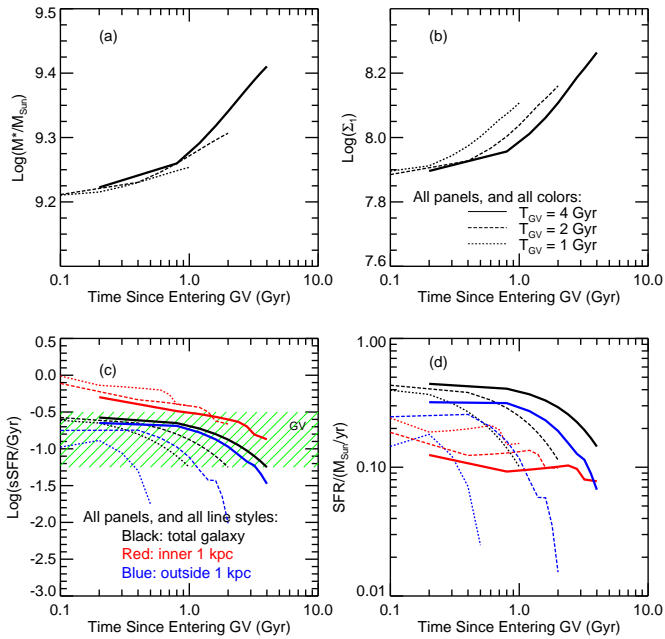}

\caption[]{Growth of a galaxy in GV in our simple models (see Section 4.1). The
four panels show (a) \mstar, (b) \sigmaone, (c) sSFR, and (d) SFR,
respectively. In each panel, different line styles show assumptions of
different time the galaxy spends in GV, as indicated in Panel (b). In Panel (c)
and (d), we also show the curves for the regions within (and outside) the
central 1 kpc in red (and blue).  The green hatched region in Panel (c) shows
the defined GV.

\label{fig:gv}}
\end{figure}

For each time step of 0.1 Gyr in GV, based on the galaxy's current \mstar\ and
sSFR, we calculate its \mstar\ and sSFR for the next step. From its new \mstar,
we infer its new \sigmaone\ from Panel (b) of Figure \ref{fig:lowzstat} (or
\ref{fig:lowzstat_delta}). Based on the change in \mstar\ and \sigmaone\ in
this step, we calculate the SFR (and sSFR) within and outside the central 1
kpc. We repeat this process until the galaxy exits GV. The results are shown in
Figure \ref{fig:gv}.

We use two conditions to constrain $T_{GV}$: (1) SFR within central 1 kpc
should not exceed the total SFR and (2) the difference between sSFRs within and
outside central 1 kpc should match the observed sSFR gradient in
\citet{liufs18}, which has a difference of about 0.3 dex within and outside the
central 1 kpc (see their Figure 5). The first condition indicates that a small
$T_{GV}$ is unlikely.  In the case of $T_{GV}$= 1 Gyr (dotted lines in Figure
\ref{fig:gv}), the central SFR exceeds the total SFR in a few hundred million
years (compare red and black dotted lines in Panel (d)). The short timescale
requires an unrealistically high SFR in the center to boost \sigmaone. The case
of $T_{GV}$= 2 Gyr meets the first condition, but fails the second one. The
difference between the central and outer sSFR is $\gtrsim$0.5 dex (red and blue
dashed lines in Panel (c)), which is larger than the observed sSFR difference
in \citet{liufs18}. The case of $T_{GV}$=4 Gyr meets both conditions and is
therefore most likely to be true.  $T_{GV}$ longer than 4 Gyr would produce a
much smaller difference between central and outer sSFRs, inconsistent with the
observed sSFR gradient in \citet{liufs18}.

Our (rough) estimate of $T_{GV}$ suggests that the time a low-mass galaxy
spending on GV is about a few gigayears. If we define this time as the
quenching timescale (i.e., time for transitioning from star forming to
quenched), our value here is consistent with the quenching timescale estimated
by using the dynamics and spatial distribution of quenched low-mass galaxies in
large halos in \citet{ycguo17dwarfenv}. It is also consistent with the estimate
of the quenching timescale of galaxies around $10^{9.5} M_\odot$ in
\citet{balogh16} and \citet{fossati17}. \citet{ji18} also estimated a quenching
timescale of a few gigayears ($\sim$4 Gyr) by using the redshift evolution of
the clustering strength of CANDELS galaxies. Similarly, \citet{phillipps19}
found timescales for galaxies crossing GV at lower redshifts $0.1 < z < 0.2$ is
also a few gigayears (2-4 Gyr). Overall, the quenching of low-mass galaxies is
not rapid, but rather a gradual and long processes. The quenching timescale can
be used to test quenching mechanisms, which need to, not only reduce the total
and outskirt SFR by a factor of 3 in a few gigayears, but also maintain the
central (within 1 kpc) SFR almost constant over this long period (see solid
lines in Panel (d) of Figure \ref{fig:gv}).

The long quenching timescale of low-mass galaxies is consistent with the low
quenching efficiency measured in many studies. For example, \citet{nancy17},
\citet{papovich18}, and \citet{chartab20} all found that the environmental
quenching efficiency decreases with the decrease of \mstar. The mass quenching
efficiency also decreases with the decrease of \mstar\
\citep[e.g.,][]{chartab20}. Both efficiencies are quite low at
\mstar$<10^{9.5}$\msun\ \citep{nancy17,chartab20,contini20}. Therefore, the
quenching timescale is expected to be long. 

One model to explain the quenching of low-mass galaxies involves both
``starvation'' (i.e., cutting off gas supplies by environmental effects
\citep[e.g.,][]{larson80,balogh97}) and ``overconsumption'' (i.e., gas
depletion because of star formation and star-formation driven outflows
\citep{mcgee14}). The gas depletion time is shorter at high redshifts when SFR
is higher, but increases with the cosmic time. In fact, the gas depletion time
for GV galaxies at $z<1.0$ is about a few gigayears \citep[inferred
from][]{genzel15}, consistent with our estimate of the quenching timescale. In
a recent paper, \citet{trussler20} found that quenching is driven by both
starvation and outflow, and they estimated the quenching timescale is a few to
several gigayears (depending on the models) in the local universe.
\citet{moutard18}, however, argued for a delayed-then-rapid quenching scenario
\citep{wetzel13}, where quenching of low-mass galaxies is rapid (a few hundred
megayears) after a long period since entering high-density regions.

The long quenching timescale also allows GV galaxies to transform their
morphology while quenching their star formation in gradual processes.
\citet{nancy17} found that the morphologies of lower-mass QGs are inconsistent
with those expected of recently quenched SFGs. Our results are consistent with
their conclusion, as both \sigmaone\ and the \sersic\ index ($n$, although only
mildly) increase when low-mass galaxies evolve from SF to GV, and eventually to
Q, as shown in Figure \ref{fig:lowzstat}. \citet{nancy17} argued that simple
gas removal processes, e.g., strangulation and ram pressure, are not able to
transform morphology and therefore some dynamical processes need to be
involved, e.g., galaxy interactions, tidal stripping, and disk fading. Our
results, however, suggest that the long star formation duration ($\sim 4$ Gyr)
in central region is able to increase \sigmaone\ and $n$. 

\subsection{Low-mass Galaxies after Green Valley} \label{discussion:after}

Once a low-mass galaxy exits GV, its total sSFR drops to at least -1.25 dex
lower than the star-forming main sequence, and is therefore too low to form
many new stars to effectively increase its total \mstar. Its \sigmaone,
however, still increases.  For example, when the galaxy we discuss in Figure
\ref{fig:gv} exits GV after 4 Gyr, its \mstar=$10^{9.4}$\msun, and its
Log(\sigmaone) is 8.25 (solid black lines in Panel (a) and (b)). From the blue
curve of Panel (b) of Figure \ref{fig:lowzstat}, we know that galaxies with
similar \mstar\ would have Log(\sigmaone/\mstar) of -0.95 in the end of Q (the
blue curve at $\Delta$Log(sSFR)$<-2$), corresponding to a \sigmaone\ of 8.45.
Therefore, its \sigmaone\ increases by 0.2 dex in Q. Since few new stars are
formed in this period, the galaxy cannot increase its \sigmaone\ by star
formation.

Therefore, \sigmaone\ has to be increased by some environmental effects through
the redistribution of existing stars. Most effects, e.g., harassment and tidal
stripping, tend to make satellite galaxies more extended rather than more
concentrated while reducing their overall \mstar\
\citep[e.g.,][]{carleton19,tremmel19}. A scenario, however, is possible to
effectively ``increase'' \sigmaone: if harassment or tidal stripping only
remove outskirt stars of a galaxy while keeping its center intact, the galaxy
will keep the same \sigmaone\ but move to lower \mstar, equivalent to
increasing its \sigmaone/\mstar\ \citep[see][and references therein]{woo17}. In
our example above, to match the observed Log(\sigmaone/\mstar)$\sim$-0.95 at
$\Delta$Log(sSFR)$<$-3 in Q, the galaxy needs to lose about 30\%-40\%
($\sim$0.2 dex) of its \mstar\ after exiting GV.

The mass loss of satellite halos due to environmental effects is well studied.
For example, \citet{jiang16} showed that a $10^{11}$\msun\ halo orbiting a
$10^{13}$\msun\ halo loses half its mass in about 2.5 Gyr. \citet{lee18} also
show that if subject to tidal stripping of a massive neighboring halo, halos
around $10^{11}$\msun\ often lose a large fraction ($> 20\%$) of their peak
mass. The fraction of the lost halo mass seems to match the required \mstar\
loss discussed above, but satellite galaxies lose halo mass more easily than
losing \mstar. As shown in \citet{engler21}, the \mstar--\mhalo\ relation of
satellite galaxies is higher than that of central galaxies. Therefore, once
becoming a satellite, a galaxy would lose a much higher fraction of its initial
halo mass than the fraction of its initial \mstar. As inferred from
\citet{errani18}, the loss of $\sim$50\% of halo mass is only corresponding to
the loss of a few percents of \mstar, much lower than the needed 30\%--40\%
discussed above.  In order to lose 30\%--40\% of \mstar, satellite halos need
to lose $\gtrsim 90\%$ of their halo mass, which seems too high to be true.

\subsection{Progenitor Bias} 
\label{discussion:bias}

One alternative explanation of the ``increase'' of \sigmaone\ in Q is
progenitor bias. It is used to explain some observed redshift evolution of
galaxy properties, e.g., the size of massive QGs
\citep[e.g.,][]{belli15,keating15,shankar15} and  the increase of massive QGs'
\sigmaone\ with redshift \citep{tacchella17}. In the redshift evolution of QGs'
size--mass relation, newly quenched systems have larger sizes and therefore
increase the average size of the quenched pool toward low redshifts.

Similarly, progenitor bias suggests that low-mass galaxies with the highest
\sigmaone\ at low redshifts are quenched at higher redshifts, when galaxies
have higher surface mass densities. These earlier quenched systems increases
the average \sigmaone\ of the quenched pool at low redshifts when comparing it
to low-redshift SFGs. This explanation is possible, as high-redshift QGs have
higher \sigmaone\ than low-redshift QGs at the same \mstar\ (Figure
\ref{fig:stat}). It, however, has another requirement: high-redshift low-mass
QGs need to have higher \sigmaone\ than their SFG counterparts. Low-mass SFGs
(down to $10^{9}$\msun) in our highest-redshift bin $1.5 \leq z < 2.5$ (where
our sample is still mass-complete, as shown in Figure \ref{fig:sample}) have a
lower \sigmaone\ than QGs in our lowest-redshift bin ($0.5 \leq z < 1.0$).
Therefore, if low-mass QGs at $z\sim2$ have a similar average \sigmaone\ as
low-mass SFGs at $z\sim2$, they are not able to explain the higher end of
\sigmaone\ of the low-redshift low-mass QGs.  High-redshift low-mass QGs need
to have a $\sim$0.4-0.5 dex higher \sigmaone\ than their SFGs counterparts to
validate the explanation.

There are two ways to satisfy this requirement. First, only SFGs with the
highest \sigmaone\ are able to quench. At \mstar$\sim$$10^{9}$\msun, the
highest Log(\sigmaone) of SFGs at $z\sim2$ is $>$8.5 (see the shaded panels in
Figure \ref{fig:ssfrvssigma1}). This value is high enough to explain the
high-end of \sigmaone\ around \mstar$\sim$$10^{9}$\msun\ at $z<1.0$. This way
implies that high \sigmaone\ at high redshifts is a direct indicator of
quenching and therefore that internal quenching could be important for low-mass
galaxies at high redshifts. The second way is that high-redshift low-mass
galaxies also undergo the dramatic growth of \sigmaone\ in GV, similar to what
we discussed in Section \ref{discussion:gv}. The second way is preferred,
because we do see a range of \sigmaone\ in high-redshift low-mass QGs from
Figure \ref{fig:ssfrvssigma1} (even with a mass-incomplete sample), which is
inconsistent with the assumption of the first way that only the low-mass SFGs
with the highest \sigmaone\ at $z\sim2$ are able to quench.

A few further observations can test the progenitor bias in low-mass galaxies.
(1) A mass-complete sample at high redshift and low mass is needed to test the
difference in \sigmaone\ between SFGs and QGs. (2) The number density of QGs at
high redshifts (and with high \sigmaone) should match that of QGs with the
highest \sigmaone\ in low redshift. (3) \sigmaone should correlates with the
age of stellar population of QGs. Namely, the higher the \sigmaone, the older
the QGs.

\subsection{Similarity and Difference between Low-mass and Massive Galaxies}
\label{discussion:diff}

An interesting and intriguing result of our study is the similarity between
low-mass ($\lesssim 10^{9.5}M_\odot$) and massive ($\gtrsim 10^{10.4}M_\odot$)
galaxies at $0.5 \leq z < 1.0$ in the relation between $\Delta$Log(\sigmaone)
(or $\Delta$Log(\sigmaone/\mstar)) and $\Delta$Log(sSFR) (see Panel (a) and (b)
of Figure \ref{fig:lowzstat_delta}). This similarity is especially true for the
dramatic increase of \sigmaone\ in GV.

As discussed in Section \ref{intro}, \sigmaone\ has been shown to be an
effective indicator of internal quenching of massive galaxies\footnote{See,
however, \citet{lilly16} for a different explanation of \sigmaone. They argued
that the higher \sigmaone\ is a consequence rather than the cause of quenching.
See also \citet{chen20} for more discussions of this explanation.}, possibly
correlated with supermassive black hole mass. \citet{chen20} found that for
massive galaxies, most black hole mass growth takes place in GV. If low-mass
galaxies are a scale-down of massive galaxies, the physical processes that
build up black hole mass and \sigmaone\ and that provide feedback from high
\sigmaone\ in massive galaxies could also work in low-mass galaxies. 

In fact, recent studies found evidence of a central engine to provide feedback
in low-mass galaxies. \citet{bradford18} found deficiencies of HI gas in
isolated AGN-host low-mass galaxies, suggesting that black hole feedback or
shocks from extreme starburst destroy or consume the cold gas. Stellar feedback
(related with central starburst) has been considered to be a major effect for
low-mass galaxies, since the fraction of AGN hosts decreases with \mstar, and
so does the importance of black hole/AGN feedback
\citep[see][though]{kaviraj19}.  However, a large sample of low-mass galaxies
have been found to have signatures of black holes \citep{reines13,reines15},
providing the sources of the central engines of quenching. \citet{greene20}
suggested that a high fraction of $10^9-10^{10} M_\odot$ galaxies hosting
intermediate-mass black holes.  \citet{penny18} found evidence for AGN feedback
in a subset of 69 quenched low-mass galaxies (in groups) in the MANGA survey.
Their result demonstrates the importance of AGN feedback in maintaining
quiescence of low-mass galaxies. 

A few differences, however, exist between low-mass and massive galaxies. First,
as we discussed in Introduction, the physical meaning of \sigmaone\ may be
different for the two populations. In massive galaxies, \sigmaone\ samples well
the core mass surface density, while in low-mass galaxies, it represents more
the inner density within a large portion of the effective radius. Second,
low-mass QGs do not have a classical bulge, while massive galaxies do. This
difference is evident by the \sersic\ index of both populations ($n \sim 2$ for
low-mass QGs and $n \sim 4$ for massive QGs, as shown in Panel (d) of Figure
\ref{fig:lowzstat}).  Last, low-mass SFGs and QGs are found to have different
projected distances to massive galaxies \citep{ycguo17dwarfenv}, indicating
different impacts of environmental effects. 

The role of the increased \sigmaone\ in low-mass galaxies could be a
``preprocessing'' of quenching. \citet{socolovsky19} found that in dense
environments at $z < 1$, compact low-mass SFGs (which would have a higher
\sigmaone\ at a given \mstar) are preferentially quenched. They proposed that
both strong feedback and environmental effects act together to quickly quench
them. By stacking spectra of high-sSFR galaxies, they even found that more
compact galaxies are more likely to host outflows. Their results are broadly
consistent with ours, both indicating that low-mass galaxies with high
\sigmaone\ are more easily to be quenched than those with a low \sigmaone\ when
they fall into a dense environment. \citet{ycguo16bursty} showed that the star
formation history of low-mass galaxies is bursty at $z\gtrsim1$. Star formation
can {\it temporarily} quench these galaxies, but new gas accretion and
recycling induce new episodes of starburst. To fully quench them, some effects,
potentially environmental, are required to turn off the continuous gas supply.

\section{Summary}
\label{summary}

We use CANDELS data to study the relationship between the central stellar mass
surface density (\sigmaone) and quenching of low-mass galaxies at $0.5 \leq z <
1.5$, 
Our mass-complete sample allows us to investigate galaxies down to $\sim
10^9$\msun\ at $0.5 \leq z < 1.0$ with reliable \sigmaone\ measurements. At a
given redshift and \mstar\ bin, we compare the mean \sigmaone\ of QGs and SFGs.
We find that for massive (\mstar\ $>10^{10}$\msun) galaxies, QGs have
significantly higher average \sigmaone\ than SFGs, consistent with other
studies in the literature.  Intriguingly, we find that low-mass QGs also have
a higher \sigmaone\ than low-mass SFGs at $z\sim1$. At $z \leq 1.0$, the
difference of \sigmaone\ between QGs and SFGs increases slightly with \mstar\
at $\lesssim 10^{10}$\msun\ and then decreases with \mstar\ for more massive
galaxies.

We use $\Delta$Log(sSFR), namely, the difference between galaxies' Log(sSFR)
and that of the star-forming main sequence at the same $z$ and \mstar, to
further divide galaxies into three categories: SF, GV, and Q. At $0.5 \leq z <
1.0$, we find that the \sigmaone\ of galaxies increases dramatically in GV, by
about 0.25 dex, regardless of their \mstar. In Q, massive galaxies reach a
plateau in \sigmaone, which correlates with (almost) no change of size and $n$.
Low-mass galaxies' increasing \sigmaone\ in Q correlates more with a gradual
increase of $n$, while intermediate-mass galaxies' increasing \sigmaone\ in Q
correlates more with a dramatic decrease of size.

We use the growth of \mstar\ and \sigmaone\ of low-mass galaxies in GV to
constrain their time spent in GV. By matching the observed sSFR gradient in
\citet{liufs18}, we estimate that the quenching timescale of low-mass galaxies
is a few ($\sim4$) Gyr at $z<1.0$. Quenching mechanisms need to gradually
quench star formation in an outside-in way, i.e., preferentially removing gas
from outskirts of galaxies while maintaining their central star formation to
increase \sigmaone. In Q, the increase of \sigmaone\ in low-mass QGs is still
unknown, with possible explanations including dynamical processes and/or
progenitor bias.

We also discuss the similarity and difference between low-mass and massive
galaxies. Although environmental effects are believed to dominate the quenching
of low-mass galaxies, the similar trend of $\Delta$Log(\sigmaone) vs.
$\Delta$Log(sSFR) suggests that internal processes in quenching low-mass
galaxies should also be investigated. Future studies can improve our
understanding in a few aspects: (1) using a mass-complete sample to explore
low-mass galaxies at higher redshifts and (2) investigating black holes/AGN in
low-mass galaxies and their relation to galaxy properties (e.g., central mass
density) in all environments.


\

The authors thank the anonymous referee for providing valuable comments, which
improved the manuscript. Support for Program number HST-AR-13891, HST-AR-15025,
and HST-GO-12060 were provided by NASA through a grant from the Space Telescope
Science Institute, which is operated by the Association of Universities for
Research in Astronomy, Incorporated, under NASA contract NAS5-26555. Members of
the CANDELS team at UCSC acknowledge support from NASA HST grant GO-12060.10-A
and from NSF grants AST-0808133 and AST-1615730. SL acknowledges the support
from the National Research Foundation of Korea (NRF) grant
(2020R1I1A1A01060310) funded by the Korean government (MSIP).

{

\appendix

\section{Robustness of \sigmaone\ Measurement in Low-mass Galaxies}
\label{app:hudf}

\begin{figure*}[htbp]
\includegraphics[scale=0.23, angle=0]{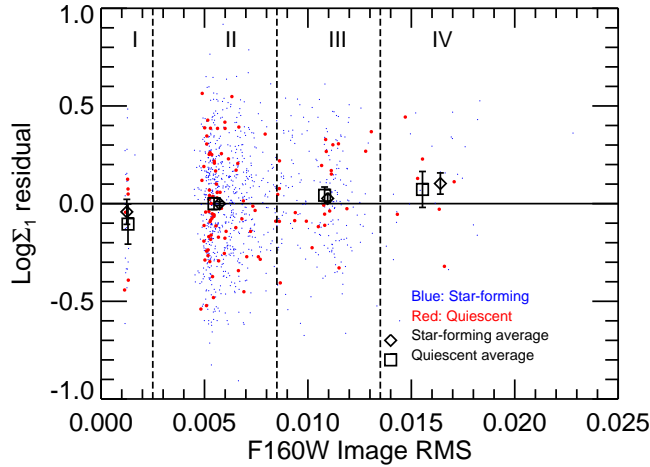}

\caption[]{Same as Figure \ref{fig:hudftest} but only using galaxies with
$8.5<{\rm log(M_*/M_\odot)}<9.5$ at $0.5<z<1.0$.

\label{fig:app_hudftest}} \end{figure*}

To further investigate the robustness of our \sigmaone\ measurement in low-mass
galaxies, we repeat our test of \sigmaone\ deviation as a function of the depth
of GOODS-S image (Figure \ref{fig:hudftest}) using only low-mass galaxies
($8.5<{\rm log(M_*/M_\odot)}<9.5$) at $0.5<z<1.0$. The result, shown in Figure
\ref{fig:app_hudftest}, demonstrates that our \sigmaone\ measurement is
unbiased for low-mass (and therefore relatively faint) galaxies in the Deep and
Shallow regions. Small deviations exists in the regions of HUDF (Region I) and
Edge (Region IV), where the numbers of galaxies are quite low. What is
important in the result is that the two populations (QGs and SFGs) have very
similar deviations (if there is any) from zero. This similarity indicates that
the small deviations do not introduce a significant systematical difference
{\it between} the two populations. Therefore, our \sigmaone\ measurement is not
only unbiased for the statistics of the whole sample (as in Figure
\ref{fig:hudftest}), but also introducing no bias to the two populations in the
low-mass sample.

\section{Effect of PSF Correction}
\label{app:psf}

It is important to test if our observed trend of the difference of \sigmaone\
between QGs and SFGs is driven by the PSF correction. To this purpose, in
Figure \ref{fig:app_corr}, we plot the PSF correction factor, namely, the
corrected ${\rm Log\Sigma_1}$ minus the raw, uncorrected ${\rm Log\Sigma_1}$,
as a function of the size (semimajor axis) of galaxies. For low-mass galaxies
(${\rm Log(M_*/M_\odot)<10}$) at $z<1$, the average correction is about 0.2
dex, with QGs (red in the figure) having a slightly larger correction factor
than SFGs (black in the figure). The difference between the correction factors
of the two populations is around 0.05 dex. This difference is much smaller than
the observed difference between the two populations (e.g., in Figure
\ref{fig:ssfrvssigma1} and Panel (b) of Figure \ref{fig:stat}). Therefore, our
main result of low-mass QGs having a higher \sigmaone\ than low-mass SFGs is
not driven by the PSF correction.

To demonstrate this point directly, we compare the raw \sigmaone\ of the two
populations in Figure \ref{fig:app_ssfrvssigma1}. This figure shows that even
without PSF correction, the raw data already display a statistically
significant difference between the two populations (i.e., QGs have a higher
\sigmaone\ than SFGs at low mass). The statistics of the difference is plotted
in Figure \ref{fig:app_stat}. Comparing it to Figure \ref{fig:stat}, we find
that the PSF correction only slightly enhances the difference between QGs and
SFGs. This enhancement is reasonable and necessary, because low-mass QGs,
which have a slightly higher \sersic\ index than low-mass SFGs, are more
affected by the PSF and therefore need larger correction factors. Moreover,
Figure \ref{fig:app_stat} shows that the trend of $\Delta \Sigma_1$ vs. ${\rm
Log(M_*)}$ is quite similar to that in Figure \ref{fig:stat}.

In principle, the PSF correction should also depend on the size of galaxies. We
argue that, however, size is only the secondary parameter in the PSF
correction. In fact, in Figure \ref{fig:app_corr}, the correction factor
already shows a dependence on galaxies' sizes, even though we do not include
size dependence in our correction method. This result is due to the coupling of
the size and the \sersic\ index $n$ in the \sersic\ profiles. Our argument is
consistent with the result of \citet{trujillo01a}, who found that $n$ is the
parameter most affected by seeing. In another paper, \citet{trujillo01b} showed
that when the effective radius of galaxies is comparable or larger than the
FWHM of the PSF, the effect of the PSF on the mean effective
surface brightness (a good proximity to our \sigmaone for low-mass galaxies) is
almost independent of galaxies' radii. Therefore, we believe that not including
size dependence in our PSF correction does not significantly affect our
results. Moreover, since smaller galaxies (e.g., low-mass QGs) are more
affected by the PSF than larger galaxies (e.g., low-mass SFGs), if we included
the size effect in PSF correction, the correction factor of low-mass QGs (or
SFGs) would have been larger (or smaller) than what we used in the paper. As a
result, the difference between the \sigmaone\ of the two populations would be
even larger, strengthening our conclusion of low-mass QGs having a higher
\sigmaone. For example, for galaxies with ${\rm 9.6<Log(M_*/M_\odot)<10}$ at
$z<1$, if we considered the size dependence in the PSF correction, the
difference of \sigmaone\ between QGs and SFGs would be larger than our reported
value by $\sim$0.11 dex. For lower-mass (or more massive) galaxies at $z<1$,
the extra difference of \sigmaone\ between QGs and SFGs introduced by
considering size dependence is smaller (or larger). 

\begin{figure*}[htbp] 
\includegraphics[scale=0.23, angle=0]{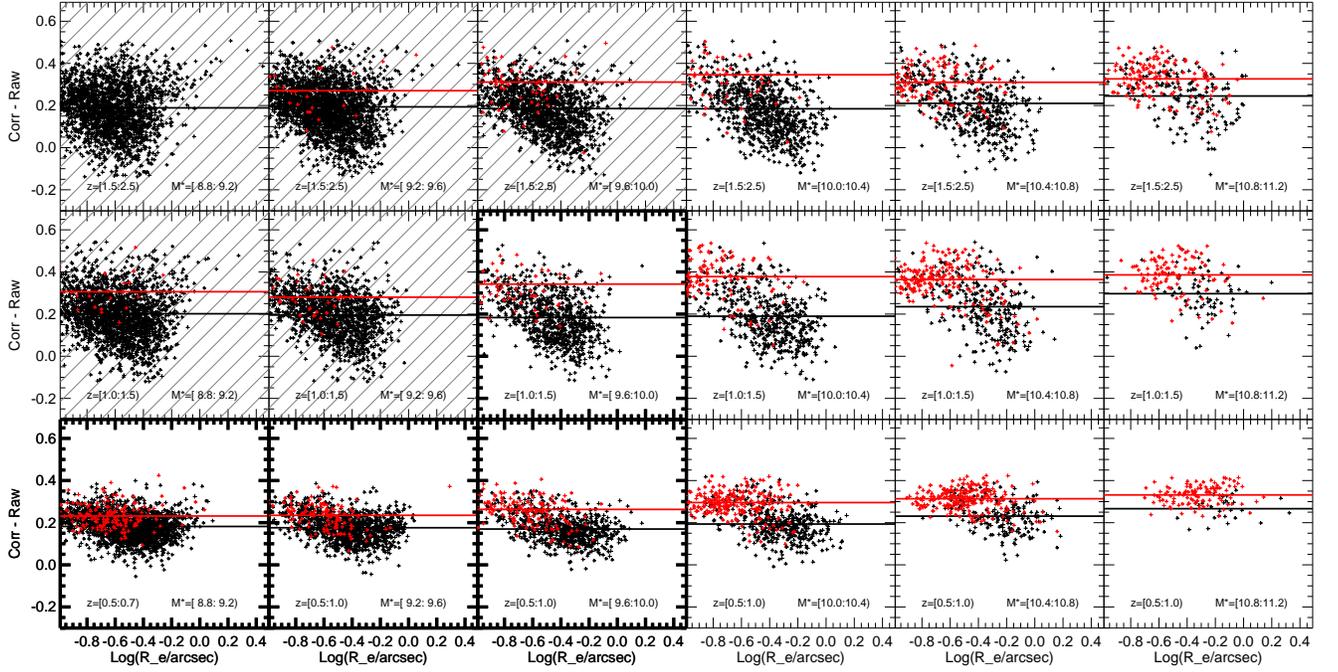}

\caption[]{PSF correction factor as a function of the size (semimajor axis) of
the galaxies in our sample. The correction factor (Y-axis) is the difference
between the PSF-corrected ${\rm Log\Sigma_1}$ and the uncorrected (raw) ${\rm
Log\Sigma_1}$. The symbols and color schemes are the same as in Figure
\ref{fig:ssfrvssigma1}. The red and black horizontal lines show the average of
the QGs and SFGs, respectively.
 
\label{fig:app_corr}}
\end{figure*}

\begin{figure*}[h!] 
\includegraphics[scale=0.23, angle=0]{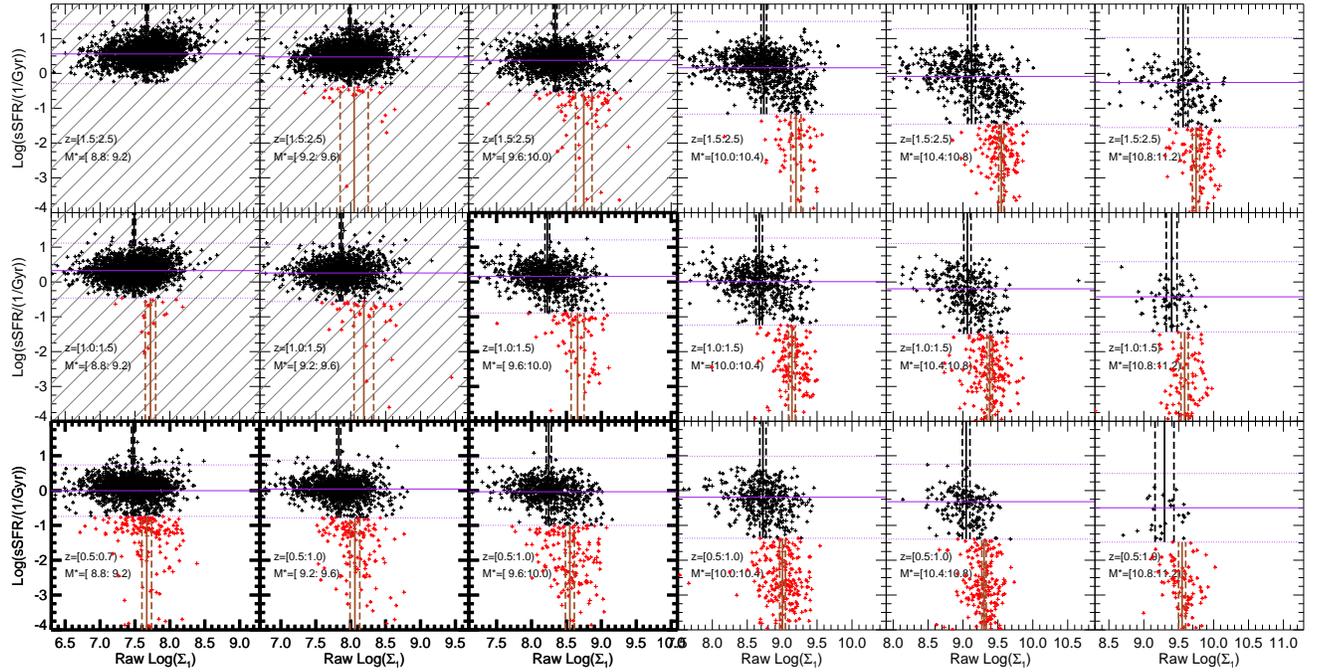}

\caption[]{Same as Figure \ref{fig:ssfrvssigma1} but using the raw \sigmaone\
(uncorrected for PSF effects). The difference of \sigmaone\ between QGs and
SFGs is already statistically significant in this case.

\label{fig:app_ssfrvssigma1}}
\end{figure*}

%
%

\begin{figure}[h!] 
\hspace*{-0.3cm}\includegraphics[scale=0.12,
angle=0]{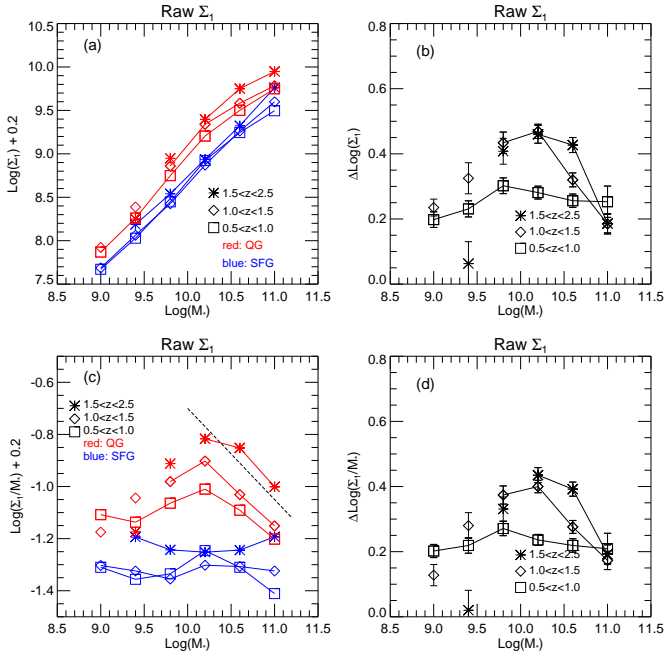}

\caption[]{Same as Figure \ref{fig:stat} but using the raw \sigmaone\
(uncorrected for PSF effects). Data in Panel (a) and (c) are arbitrarily moved
up by 0.2 dex to keep the same Y-axis range as in Figure \ref{fig:stat} for a
direct comparison of the slopes.

\label{fig:app_stat}} 
\end{figure}

}


\end{document}